\documentclass[10pt]{article}

\usepackage{ucs}
\usepackage[utf8]{inputenc}

\usepackage{url,hyperref,fullpage}

\usepackage[OT2,OT1]{fontenc}
\def\cyr{\fontencoding{OT2}\fontfamily{wncyr}\selectfont}

\newtheorem{definition}{Definition}
\newtheorem{theorem}{Theorem}
\newtheorem{corr}{Corollary}

\newcommand{\sdom}{\mbox{\small \rm dom~}}
\newcommand{\dom}{\mbox{\rm dom~}}

\title{Zeno machines and hypercomputation}

\author{Petrus H. Potgieter\\
Department of Decision Sciences, University of South Africa\\ 
PO Box 392, Unisa 0003, Pretoria, {\tt potgiph@unisa.ac.za}.
}

\begin{document}

\maketitle

\begin{abstract}
This paper reviews the Church-Turing Thesis (or rather, theses) with reference to their origin and application and considers some models of  ``hypercomputation'', concentrating on perhaps the most straight-forward option: Zeno machines (Turing machines with accelerating clock). The halting problem is briefly discussed in a general context and the suggestion that it is an inevitable companion of any reasonable computational model is emphasised. It is suggested that claims to have ``broken the Turing barrier'' could be toned down and that the important and well-founded rôle of Turing computability in the mathematical sciences stands unchallenged.

\end{abstract}

\section{Introduction}

The popular and scientific literature in foundational aspects of computer science and in physical science have of late been replete with obituaries for the Church-Turing Thesis and triumphant announcements of the dawn of ``hypercomputation'' (the ``scare quotes'' around which are to be carelessly omitted from this point on). Some of the proponents of this idea and their readers believe that somehow the Church-Turing Thesis has been disproved. It is often not quite clear however what exactly they take the Thesis to mean. Indeed, some authors even claim Turing's oracle machine (\emph{O-machine}) to have been a precursor of a new kind of computation, which obviously it was not \cite{HodgesOnCopeland}. This paper is an attempt to identify the nature of the claims made by several hypercomputing models and to make clear to what extent they do---and do not---challenge the established notion of computability.

Mathematicians would note that claims to hypercompute call for a re-examina\-tion of the quite well-known solutions of the \emph{Entscheidungsproblem} as well as Hilbert's tenth problem and should be considered very critically. In computer science, applied mathematics and operations research these claims could challenge the formal definitions of \emph{algorithm} or \emph{computation} as they have now been used for more than half a century. We start by reviewing the Church-Turing Thesis in its various forms, and proceed to give an overview of some unusual or novel computational schemes.

\section{Formal computation}

The necessity of a formal definition of \emph{algorithm} or \emph{computation} became clear to the 
scientific community of the early twentieth century due mainly to two problems posed by David Hilbert:
\begin{itemize}
\item the \emph{Entscheidungsproblem}---can an algorithm be found that, given a statement in 
first-order logic (for example, in Peano arithmetic), determines whether it is true in all models of a 
theory; and 
\item Hilbert's tenth problem---does there exist an algorithm which, given a Diophantine equation, 
determines whether is has any integer solutions?
\end{itemize}
The \emph{Entscheidungsproblem}, or \textit{decision problem}, can be traced back in some form at least to Leibniz and was 
successfully and independently resolved in the negative during the mid-1930s by Alonzo Church \cite{church1936} and Alan 
Turing \cite{turing1935}. Church defined an \emph{algorithm} to be identical 
to the definition of a function in his lambda calculus. Turing in turn first identified 
an algorithm to be identical with a \emph{recursive function} and later with a function 
computed by what is now known as a \emph{Turing machine}. It was shown that the recursive 
functions, Turing machines, and the lambda calculus define identical classes of functions. The 
remarkable equivalence of these three definitions of disparate origin strongly supports  
the idea of this as an adequate definition of computability and the \emph{Entscheidungsproblem} 
is generally considered to be solved. 
Using the same definition of computability Yuri Matiyasevich 
showed \cite{matijasevich1970ru} in 1970 (refining and completing work by Julia Robinson, Martin Davis and Hilary Putnam) 
that, within the Church-Turing framework of 
computation, no algorithm for determining whether a given Diophantine equation admits 
integer solutions exists. 

The negative answers to the \emph{Entscheidungsproblem} and to Hilbert's tenth problem are related 
to the \emph{halting problem} for Turing machines: given a program or input for an arbitrary Turing machine together with a description of the machine, 
is it possible to determine algorithmically (i.e. using some other, fixed, Turing machine) whether the 
first Turing machine ever halts on the given program or input? The answer to this question 
is NO. The function $f_T$ which assigns to each $x$ the value 1 if the Turing machine $T$ halts 
on input $x$ and 0 otherwise is the classical example of a non-computable function for a 
sufficiently powerful $T$, specifically when $T$ is a \emph{Universal Turing Machine} (UTM) which is 
capable of simulating all other Turing machines.

Unless it is indicated or directly implied otherwise, ``computation'' and ``compute'' in the paper will refer to the action of and functions computed by an appropriate Turing machine.

\section{The Church-Turing theses}

There are many versions and interpretations of what is loosely known as the Church-Turing Thesis, 
and it would naturally be more accurate to speak in the plural, of Church-Turing \emph{theses}. 
The equivalence of the \textit{definitions} of computability given by Church and by Turing gave 
rise to the \emph{Church-Turing Thesis} (CTT) which, as formulated by Turing, states that 
\begin{quote}
Every `function which would naturally be regarded as computable' can be computed 
by a Turing machine. \hfill (CTT)
\end{quote}
Now, certainly any recursive function or function defined by lambda calculus or by any of a number 
of other computational schemes, including Markov algorithms, can \emph{provably} be 
computed by a Turing machine. The problem is that the vagueness of the concept 
``function which would naturally be regarded as computable'' 
ensures that the statement of the CTT above is not something that can ever be 
proved. The following remarks (as they appear in \cite{stanford-encyc}) by Alan Turing will shed some 
light on \textit{his} views of the universal machine and of `natural' computability.\\
\begin{quote}
A man provided with paper, pencil, and rubber, and subject to strict 
discipline, is in effect a universal machine. \hfill (1948) \\
\end{quote}
\begin{quote}
The idea behind digital computers may be explained by saying that 
these machines are intended to carry out any operations which could 
be done by a human computer. \hfill (1950) \\
\end{quote}
\begin{quote}
The class of problems capable of solution by the machine [the Automatic Computing Engine (ACE)] 
can be defined fairly specifically. They are [a subset of] those 
problems which can be solved by human clerical labour, working to 
fixed rules, and without understanding. \hfill (1946) \\
\end{quote}
\begin{quote}
Computers always spend just as long in writing numbers down and 
deciding what to do next as they do in actual multiplications, and it 
is just the same with ACE ... [T]he ACE will do the work of about 
10,000 computers \dots Computers will still be employed on small 
calculations \ldots \hfill (1947)\\
\end{quote}
An absolutely crucial point here is that Turing was thinking about a \emph{human 
computer} which, in his time, was a calculating clerk. CTT as stated above and with Turing's sense of `naturally computable' is what should be referred to as \emph{the} Church-Turing Thesis and it is the only very well-established one of the four versions in this section. 
A stronger version of CTT, call it the Physical CTT (PCTT) would be
\begin{quote}
Every function that can be physically computed, can by computed by a Turing machine. \hfill (PCTT)
\end{quote}
Here the notion ``physically computed'', if including the result of any physical process (or of 
our model of any physical process), is so vague that the thesis PCTT cannot but be false. For example, 
a Turing machine cannot effectively approximate any of the values of one-dimensional Brownian 
motion at rational points in time almost surely (with respect to Wiener measure), as discovered  
by Willem Fouch\'e \cite{foucheJSL2000} recently. 
\emph{Gandy's Thesis} \cite{stanford-encyc} (or, as Gandy called it, the \emph{Thesis M}) posits that 
\begin{quote}
Whatever can be calculated by a machine (working on finite data in accordance with a finite program of instructions) is Turing-machine-computable.
\end{quote} 
Yet another version of the Thesis is the following, as stated in \cite{bernstein97quantum}:
\begin{quote}
Any `reasonable' model of computation can be \emph{efficiently}\footnote{
An efficient simulation is one whose running time is bounded by some polynomial in the 
running time of the simulated machine.} simulated on a probabilistic Turing machine. \hfill (SCTT)
\end{quote}
This version is the Strong Church-Turing Thesis (SCTT) and it might very well be provably false, unlike CTT and Gandy's Thesis which---this paper will argue---remain eminently reasonable suppositions. 
In this regard \cite{DavisMyth}, \cite{Akl2005}, \cite{Yao2003} and \cite{TimpsonPhD} are well worth also consulting.

\section{Zeno machines}

Hermann Weyl in 1927 raised \cite{weyl1949} the possibility of a feasible device 
performing an infinite number of steps in finite time. We should understand the steps  
to be in some sense identical except for the time taken for their execution\footnote{Incidentally, the expression of this idea precedes Turing's precise formulation of his precise ideas around computability.}. 
It can be easily 
seen that allowing this kind of speed-up of the computing device, one can solve the 
halting problem for Turing machines in finite time. The solution, however, is \emph{not} by a 
Turing machine but by what may be called a \emph{Zeno machine} (ZM) (or Accelerated Turing Machine, see  \cite{Copeland2002} or \cite{Copeland2004}; or, in \cite{BoolosJeffrey1980}, simply \textit{Zeus} by Boolos and Jeffrey).
As acknowledged elsewhere in the literature, e.g. in \cite{CaludePaun2004}, hypercomputational schemes tend to be proposals for somehow accomplishing infinitely many computational steps in a finite time, so it is apt to first focus on the ZM in a general examination of hypercomputation.

Without loss of generality, we shall assume a ZM to be identical to a Turing machine with 
one input tape, one output tape and a storage tape \emph{except} that the ZM takes $\frac{1}{2}$ 
hour to execute the first transition, $\frac{1}{4}$ hour for the second, $\frac{1}{8}$ hour for 
the third etc. After one hour the ZM will have finished its operation and one will perhaps find 
the answer to some tantalising question on the output tape. For example, a ZM which, on input 
of a TM machine description $m$ and natural number $n$, follows the instructions
\begin{quote}
{\bf begin}\\
\indent write {\bf 0} on the first position of the output tape;\\
\indent $i \leftarrow 1$;\\
\indent {\bf do} \\
\indent \indent run TM $m$ on input $n$ for $i$ steps;\\
\indent \indent if $m$ has halted, write {\bf 1} on the first position of the output tape;\\
\indent \indent $i \leftarrow i+1$;\\
\indent {\bf while} $i>0$;\\
{\bf end.}
\end{quote}
will after one hour have a {\bf 0} on the output tape if TM $m$ does not halt on input $n$ and 
a {\bf 1} if TM $m$ would indeed halt on this input. In some sense this ZM therefore solves the 
halting problem in finite time and we have observed an appropriately constructed ZM 
output a function that cannot be computed by a Turing machine. 
Unfortunately however (see \cite{svozil1998} and below) there exists no ZM that can solve the halting problem \emph{for ZMs}. 

A typical application for a ZM would be deciding L.E.J. Brouwer's question whether 777 appears in the decimal expansion of $\pi$ (which question also occupied Wittgenstein). By a similar but even more elementary scheme than the above, we can describe a ZM that outputs {\bf 1} 
if it does indeed appear and {\bf 0} otherwise. 
A more interesting situation is encountered when we try to disprove the Twin Prime Conjecture (TPC). Recall that the TPC states that there are infinitely many pairs of prime numbers of the form $(n,n+2)$. In a small flight of fancy one can imagine a ZM executing the following program:
\begin{quote}
{\bf begin}\\
\indent $i \leftarrow 1$;\\
\indent {\bf do} \\
\indent \indent write {\bf 0} on the first position of the output tape;\\
\indent \indent check for twin prime pairs from $i$ up to $i+100$;\\
\indent \indent if any were found, write {\bf 1} on the first position of the output tape;\\
\indent \indent $i \leftarrow i+100$;\\
\indent {\bf while} $i>0$;\\
{\bf end.}
\end{quote}
Now, if the execution time of the ZM is over and {\bf 0} is in the first position of the output tape, then clearly the TPC is false because from some point $i$ onward no further twin primes would have been found. But what if the TPC is true? Now, the first character on the output tape would have changed infinitely many times from {\bf 0} to {\bf 1}---and back---so the terminal state of the device is simply undefined. 
This example illustrates one of the problems with Zeno Machines: in some way we believe the terminal state of the machine to be well-defined since we need to read off some result, but how? 

Well, since the ZM is based on a TM, we might---not unreasonably---presume the state of the output tape as well as the internal state of the machine to have to be well-defined at the end of the computation. Horrors that might otherwise appear including a write head that has ``run away to infinity'', or an output tape in undefined condition (as is possible with the TPC program above) or an undefined internal state (which may happen in the case of the ZM for the halting problem). Just as in the case of Thomson's lamp\footnote{
Thomson's lamp is an electric light, the switch of which is flipped at intervals of length $\frac{1}{2}$, $\frac{1}{4}$, $\frac{1}{8}$, \ldots } one might argue that any terminal state of the device after one hour is consistent with the execution of the super-task\footnote{Thomson's term again, in \cite{Thomson1954}.
} but this means that a {\bf 0} on the tape would be consistent with an infinite number of ones having been written in that position during the course of the calculation and then we would have lost the ability to infer from the output {\bf 0} that the TPC were false. On the other hand, 
one could argue that the inference from {\bf 0} on the tape that only a finite number of pairs were found was totally wrong and unwarranted, but this again leaves one wondering about the utility of this kind of machine.

One needs to identify a terminal state of the devices (at least sometimes) in order to define the output of the machine. 
Could we simply \textit{declare} that any ZM has, in each computation, to stabilise some positive time before before one hour has elapsed in our description, i.e. after only finitely many steps? 
Fine, but then we have only the halting Turing computations and still need to fix the problematic cases. This we could do by declaring that in such cases the output tape be frozen, the write head brought back to the beginning of the tape, the internal state of the machine stabilised and a special flag set, notifying us that we should not interpret the output on the tape as justly obtained. However, what we have done now is to simply incorporate an oracle for the halting problem into the ZM! With such an oracle, who would need a Zeno Machine anyway? 

Is there any indication the a ZM could be realised physically? Well, in Malament-Hogarth space-times (which apparently include our universe), it is possible for an observer in finite time to 
observe arbitrarily long calculations due to relativistic effects. This has been described in some 
detail by Etesi and N\'emeti (\cite{etesi} and \cite{etesinemeti}). Devices operating in Malament-Hogarth space-times will still present severe problems with the definition of the final state of the (putative) computation \emph{à la} Thomson's Lamp. 
Relativistic machines can sidestep this issue somewhat by imagining that the machine or the observer is destroyed by a black hole at just the right moment, if one can consider such a gravitonic intervention as a legitimate computational step.

The rest of this section mentions two classes of devices related to the class Zeno machines.

\subsection{Inductive Turing machines}

The \textit{inductive Turing machines} were introduced by Putnam \cite{Putnam1965} and Burgin \cite{Burgin1983}. These devices operate exactly as three-tape Turing machines with the exception that the inductive Turing machine (ITM) is said to compute an output $y$ from input $x$ if the output tape contains $y$ after some finite number of steps and then no longer changes. An ITM may be said to compute a certain output even if the machine never reaches a halting state. In this view, and ITM is a ZM with the meta-halting concept that the output tape must be stable after finitely many computation steps. As with a ZM, unless the actual classical halting state has been reached, one will not in general know after a finite number of steps whether the content of the tape corresponds to the output of the machine corresponding to a give input. Therefore the output is in general only defined after having considered all the countably infinitely many computational steps possible for the machine. The internal state of the machine, and that of the work tape, might again display a disconcerting resemblance to Thomson's lamp. The sets decidable by ITMs fit nicely in the arithmetic hierarchy---they are exactly all of  $\Delta^0_2=\Sigma^0_2\cap\Pi^0_2$ \cite{BurginKlinger2004} and can solve the halting problem for Turing machines.

\subsection{Infinite time Turing machines}\label{secittm}

Infinite time Turing machines \cite{Hamkins2005} were introduced by Hamkins and extend the ordinary TM model by defining the configuration of the device at every ordinal ``time''. The operation from an ordinal to its successor is simply an ordinary TM step and the configuration of the device at a limit ordinal is defined as consisting of the $\lim\sup$ of the tape contents with the head at the beginning of the tape and the machine in an initial state. There are no logical contradictions in the infinite time Turing machine metaphor and they have given rise to an interesting theory in which, for example, $\mbox{\bf P}\neq\mbox{\bf NP}$, and $\Pi_1^1$ sets are decidable by these devices. However, there is no suggestion at all of how such devices might be engineered or even conceived in a physical theory (nor is it necessary if these devices are considered in a logic context only) so these are ``machines'' in name only. 

\section{The halting problem for Zeno machines}

The halting problem (HP), the unsolvability of which has played such an important rôle in the Hilbert programme and in the development of theoretical computer science, can be stated in a relatively general context and its unsolvability shown to follow from rather mild assumptions about the class of machine under consideration. We shall assume throughout that tokens $0$  and $1$ (usually the binary numbers themselves) are valid outputs of all the machines under consideration and that each machine $X$ defines a partial function $\psi_X$ on the input space with values in the output space. For a more general and extensive treatment of HP and several cognate notions the reader is referred to \cite{yanofsky}.

\begin{definition}
Call a class of computing devices $\mathcal C$ with uniform input and output devices a \emph{Chatrapur\footnote{The author introduces the informal definition and terminology here in view of the reported conception of Alan Turing in the city of Chatrapur, in the modern-day Indian state of Orissa. Turing's result is restated, for an arbitrary Chatrapur class.} class} of machines if
\begin{enumerate}
\item there exists a representation scheme $X\rightarrow n_X$ such that 
for each machine $X$ (computing a partial function $\psi_X$) one may use $n_X$ as input to any machine in the class $\mathcal C$;
\item whenever $Y\in \mathcal C$ computes a total function $\psi_Y$ with values in $\{0,1\}$ then there exists a machine $X\in \mathcal C$ such that $\psi_X \equiv \left.\psi_Y\right|_{\psi^{-1}_Y\left(\{0\}\right)}$.
\end{enumerate}
\end{definition}

The second condition means that $\psi_X$ is identical to $\psi_Y$ where $\psi_Y=0$ and undefined otherwise---this being  what one needs to prove the unsolvability of HP. The first condition is not very exacting either---amounting (roughly) to the the existence of  programmable machines in the class (some of them possibly universal).

\begin{definition}
The halting problem (HP) for a class of machines $\mathcal B$, with uniform input and output devices where each $X\in \mathcal B$ computes a partial function $\psi_X$, is said to be \emph{solved by $f$} if 
$$f(n_Z)=\chi_{\sdom \psi_Z}(n_Z) \quad \mbox{for each $Z\in \mathcal B$.}$$
The \emph{halting problem} proper for a class of machines $\mathcal B$ is the question whether there exists a machine $Y\in\mathcal B$ such that $\psi_Y$ solves HP for $\mathcal B$.
\end{definition}

The following theorem is simply a restatement of Turing's 1936 proof of the unsolvability of HP for Turing Machines, phrased in the terms introduced here.

\begin{theorem}
\label{th:hp}
There is no Chatrapur class in which the halting problem is solvable.
\end{theorem}

{\bf Proof. }
Assume that the machine $Y\in\mathcal C$ solves the halting problem for the Chatrapur class $\mathcal C$. Let $X$ be as in the second Chatrapur condition and consider $X(n_X)$. Now, since $\psi_Y$ is total, either $\psi_Y(n_X)=0$ or $\psi_Y(n_X)=1$. In the first case, by the fact that $Y$ solves the halting problem, $n_X \not\in \dom\psi_X$ but this is a contradiction since by definition of $X$, $\psi_X$ is defined on $\psi_Y^{-1}\left(\{0\}\right)$. If, on the other hand, $\psi_Y(n_X)=1$ then  $n_X \in \dom\psi_X$ by definition of $Y$ but since $\psi_X=\psi_Y=0$ on $\dom\psi_X$, this is likewise a contradiction.
{\hfill Q.E.D.}

It is easy to see that the Turing machines, for instance, form a Chatrapur class. In extending the discussion to Zeno machines one is compelled to introduce a halting concept but we shall attempt to place as few restrictions as possible on the operation of the machine.
\begin{itemize}
\item 
Whenever the output tape has a limit as the number of computational steps go to infinity (i.e. the content of each cell stabilises) and the position of the head has the limit $2$ (i.e. it stabilises over the second cell) when the machine receives input $n$ then we'll say that the machine \textit{computes} the limiting value in the first cell. For a (possibly partial) bit-valued function this is taken to be the only way it can be said to be computed by a ZM.
\item 
On the other hand, if a ZM on input $n$ never stabilises the first cell of the output tape, we'll take $n$ to lie outside the domain of the partial function computed by the machine (i.e. Thomson's Lamp may not start an output string, as it were). 
\end{itemize} 
These assumptions about the functions that are computed by a ZM are not extravagant. With this partial halting concept a ZM can already solve the halting problem for classical Turing Machines. However they are sufficient to ensure that no ZM for solving the halting problem for the class of ZMs exists. This follows directly from Theorem \ref{th:hp} and the following observation.

\begin{theorem}
The class $\mathcal Z$ of Zeno Machines is a Chatrapur class.
\end{theorem}

{\bf Proof. }
Satisfaction of the first Chatrapur condition is trivial since a ZM has the same description as a corresponding TM. Consider the second condition by letting $Y$ be a ZM computing a binary-valued total function $\psi_Y$ using the halting concept described above. Construct $X$ by adding a tweak to the description of $Y$: whenever $Y$ writes $1$ in the first position of the output tape, branch into a parallel (hyperactive) version of the operation of $Y$ with the only modification being that the first cell in the output tape is flipped at each step in the operation and the halting state is replaced by an infinite loop (with continued flipping of the bit in the first cell) and whenever an instruction of the original machine $Y$ writes a zero to the first cell, drop out of the hyperactive state and follow the original operation of $Y$ (where even halting normally is allowed). $X$ can now be seen to have the properties required by the second Chatrapur condition.
{\hfill Q.E.D.}

\begin{corr}
The halting problem for Zeno machines is not solvable by a Zeno machine.
\end{corr}

Since the unsolvability of HP is a direct result of the Chatrapur conditions, it becomes inevitable in any class of machines that can take their own description as input and which are powerful enough to satisfy the second condition. Solving HP for some subclass of machines but not by a machine in the subclass itself\footnote{HP for a subclass of Turing Machines is in fact solvable by TM---for example, by the TM  computing the constant function $1$.} is not necessarily interesting of itself. 
In this way, Turing's unsolvability proof encapsulates a principle which is more fundamental and interesting than the mere finding about the mechanics of Turing Machines that is is often taken to be. As put quite nicely in \cite{TimpsonPhD},
\begin{quotation}
The halting problem, then, tells us nothing about what can be built; it tells us the mathematical constraints on what can be computed given the way we have defined computing.
\end{quotation} 
which this subsection has attempted to illustrate. A similar exposition can be found in \cite{svozil1998} and in \cite{Cotogno2003}.

Ord and Kieu \cite{OrdKieu2005} have examined the halting problem in a similar context and although they do not disagree with the premises or with the conclusion, they do not find it an argument detracting from the value of hypercomputation, essentially on the basis that for many classes of proposed hypercomputation devices one either does not really want to solve the generalised halting problem, or does not want the class of devices to satisfy the Chatrapur conditions.

\section{Other non-conventional computing}

It is fascinating and delightful that a physical process can be conceived that 
solves the halting problem, for example a quantum process or a relativistic machine. 
In our view the Church-Turing Thesis, in the usual form (CTT), is quite independent of such experiments. Turing, after all, was trying to delineate mathematically the power of a \emph{human} computer when working to a finite number of clearly enunciated rules.  
Whatever one calls `computable' should be by a process which is in principle physical but the converse---that any physical process should be considered a computation---demolishes the very distinctness of the notion of computability. It has long been quite clear that certain physical processes (or, at least, their models) give rise to non-computable objects. 
It was shown in 1982 for example \cite{pourel1982} by 
Pour-El and Richards that computable boundary conditions for the three-dimensional wave equation 
exist that give rise to non-computable solutions. More recently, Fouch\'e has pointed out 
\cite{foucheJSL2000} that for a \emph{generic} one-dimensional Wiener process (one from a 
certain measure one set of trajectories) \emph{all} the values assumed by the process at recursive real numbers are non-recursive. Fouché's result really implies that if a Wiener process trajectory is selected at random then with probability one the value of the process at \textit{every} rational (and other recursive) time will be a non-computable real numbers. One can think of this as meaning that when a grain of pollen (recalling Robert Brown and confusing physical reality and its model for a moment) is observed at rational intervals, then it will always be observed at an uncomputable position in space.

The idea that a physical process can produce non-computable output (as also pointed out it \cite{stanford-encyc}) is therefore not novel. 
Quantum computing, machines exploiting relativistic time effects and bio-molecular computing are popular tools for researchers attempting to ``jump the Turing barrier'' and the rest of the section consists of a brief survey of their power and application.

\subsection{Quantum computation}

It is well known that in 1982 Richard Feynman, in a talk at MIT \cite{feynman1982}, proposed 
using quantum mechanical phenomena in order to perform computations. He identified the immense 
compactness of representing information in \emph{qubits} (\textit{qu}antum \textit{bits}) due to the superposition principle and 
the potential of quantum systems to be used in practical computation. The same ideas were in fact expressed\footnote{
In a Russian text-book \cite{Manin1980}. Also see \cite{potgieter2004}  or \cite{potgieteronmanin}, the latter with the actual text and a translation into English. Manin's remarks are well-known to Russian authors and cited in \cite{KitaevEA} for example.
} by Yuri Manin somewhat  earlier. The principle is that a finite-dimensional quantum system is somehow prepared in an initial state and acted upon by operators corresponding to normal events in quantum computation. The action of the operators transposes the state of the system to one which reveals meaningful results when the complex quantum state of the system is disturbed through measurement.  The intricacies of quantum interactions can, it is thought, give such a device some advantage over a classical computer. 
In 1985, David Deutsch \cite{deutsch85quantum} proposed 
a universal quantum computer (UQC) for this model, similar to the universal Turing machine (UTM), which should simulate any other quantum computer, i.e. which should be able to compute any function computable by a quantum Turing machine which is specified as part of the input. Deutsch's construction has been questioned  \cite{Shi2002,Myers1997} and the existence of a universal\footnote{
Universality is a substantial and important issue to be examined in a later contribution.
} quantum computer (in this sense) has not been clearly established in the eyes of many. 
Any quantum computation of this sort can be simulated  (but \emph{not} necessarily effectively) by a probabilistic Turing machine (PTM) and hence on a deterministic Turing machine (see \cite{simon94power} or \cite{bernstein97quantum}, for example, for a detailed discussion of why this is the case). 

What then are the successes of quantum computing so far? 
Quantum computing offers great potential for speeding up calculations and the 
following milestones are worth noting: 
Shor's 1994 algorithm for integer factorisation \cite{shor-introduction}; 
Grover's 1996 algorithm \cite{grover96fast} to speed up search in an unordered list, with quadratic speedup over its classical counterpart. 
A quantum Turing machine as defined in \cite{deutsch85quantum} cannot however compute any function 
that cannot also be computed by an ordinary probabilistic Turing machine. Since the quantum computer 
is apparently faster than a classical computer for several problems it would seem that the Strong 
Church-Turing Thesis \emph{is} perhaps violated by quantum computers as illustrated in 
\cite{bernstein97quantum} but as yet no example has been found that would survive a proof of $\mbox{\bf P}=\mbox{\bf NP}$. 

It is sometimes argued that because the fundamental nature of reality is quantum mechanical rather than classical, quantum computing must be more powerful than classical computing but this is a complete fallacy, as is the suggestion that since quantum cryptology is more powerful than classical cryptology (which is really the case) the same should follow for quantum computing. 
Since the classical Church-Turing Thesis (CTT)  places no restriction on the time required to simulate the computation on an ordinary Turing machine, quantum computation (via quantum circuits as in \cite{KitaevEA} or Deutsch-type quantum Turing machines) is therefore at best (possibly faster) ordinary computation using other hardware.

\subsection{Quantum processes}\label{secqp}

Quantum processes beyond the established definition of  quantum computation by QTM or by quantum gates provide some further interesting and curious results. 
In a recent publication, Calude and Pavlov \cite{calude2002} have described a quantum 
process that solves the halting problem correctly with probability tending 
to 1 constructively as the time for computation tends to infinity. Their (theoretical) 
device therefore evaluates a function which cannot be computed by a Turing machine\footnote{
A popular discussion, copied from \emph{New Scientist} magazine may be found at 
\url{http://www.cs.auckland.ac.nz/~cristian/smashandgrab.htm}.}. 
However, as clearly pointed out by the authors, their device operates in an infinite 
dimensional Hilbert space in contrast to canonical quantum computing. 
A similar notion of Nielsen is discussed in \cite{TimpsonPhD}.

T.D. Kieu has proposed \cite{Kieu2004} another interesting process: a procedure to determine whether an arbitrary Diophantine system has solutions or not, in apparent violation of the accepted solution to Hilbert's Tenth Problem. As Kieu himself states clearly, this is not a ``standard quantum computation''  (in his terminology, in \cite{Kieu2004}). 
Kieu's solution method has been roundly criticised in some corners (for example \cite{Tsirelson2001} to which Kieu \cite{Kieu2005} has replied, but probably not yet conclusively) and the terminology is rather puzzling since in the usual 
literature, a ``quantum adiabatic'' process has the same computing power as a ``standard quantum computation'' (recently proved in \cite{AharonovEA2004}). The Kieu process is, however, not one that can be simulated in the sense of Deutsch \cite{deutsch85quantum} or Bernstein and Vazirani \cite{bernstein97quantum}.

\subsection{Natural computing}\label{secnc}

By \textit{natural computing} one essentially refers to models inspired by biology. 
Adleman in the 1990s demonstrated the solution of the Hamiltonian path problem for a 
seven-node graph using DNA 
computing (see \cite{reif98alternative} or \cite{reif-emergence}). The main advantage of DNA 
computing appears to be the physical compactness, the low cost of the data storage, and the easy reproducibility of a setup for massive parallelism. The main immediate problem is that of autonomous computing, i.e. of performing more than a small number of steps without intervention from outside the computing system (by the lab assistant, for example). It also seems unlikely that a universal biomolecular computer will be conceived. DNA computing does not challenge established notions of computability. It has been suggested in \cite{CaludePaun2004} that another metaphor inspired by biology, the \textit{membrane computing} of P\u{a}un, might allow the solution of the halting problem for Turing machines in a physically realistic manner.

\section{A challenge to hypercomputation}

Clearly a computation should in principle be a physically realisable or (at least) theorisable process. 
There is a school of thought that holds that any conceivable physical or apparently physical  process may be considered to be a computation, but what then is the sense of having a definition of \emph{computation} at all? Could not, otherwise, any problem arising in practise simply be left to solve itself by physical simulation? The notion that, just because something happens in reality (or in physical theory), we should put a ``computational'' model on it seems a bit fanciful. Few people, after all, would think of Brownian motion as a ``computation''. The infinite-time Turing machines from \ref{secittm} cannot be considered candidates for hypercomputation on the grounds of there being no indication of how their operation might be realised physically.

There is also no great novelty in describing processes that are supported by classical or quantum physics that allow the observation of non-computable results \footnote{
How, actually, since we shall only ever check a finite initial segment of the output? But that is another question...
} as the results of  Fouché and of Pour-El and Richards mentioned above clearly show. 
Neither is it a breakthrough to simply conjecture a model in which interesting problems can be solved---this can be done by simple Zeno machines (apart from the problem of defining the final state).

One task of those studying hypercomputation should be to describe physically realisable processes leading to \textit{interesting} results. 
The reported ability of the quantum processes of \ref{secqp} and the membrane computing mentioned in \ref{secnc} 
to solve the halting problem or Hilbert's 10th problem is much more interesting therefore than the action of Zeno machines---which, after all, could be said to be able to solve the same problems---since the description of the process is much closer (in the absence of a conveniently situated Malament-Hogarth space-time) to what we perceive as a real and executable physical process. 
Whether these processes providing interesting ``outputs'' should be called \textit{computations} of any kind or not is then a matter of (changing the) convention, against which change a mild prejudice is freely admitted by the author for reasons given below.

\section{Conclusion}

The question whether every physically realisable or theorisable process should be considered a computation, should be taken as settled in the negative. 
A radical revision in our view of space and time (see \cite{longo2002} for a very non-technical 
overview) and/or human nature might however in the future necessitate a wide-ranging revision of our ideas about 
mechanical computability. Until such time \textit{hypercomputation} will be a useful sobriquet for physical processes that have some semblance to but are distinct from what we have properly regarded as constituting computations since Turing, Church, Post, Markov, Péter e.a. When hypercomputational schemes are introduced, one should be extremely careful as to how they are defined and how their physical aspect is formulated in order to establish that one is doing something other than introducing a partial view of the arithmetical hierarchy. 
Many of the advocates of hypercomputation are well aware of the questions raised in this contribution (for example \cite{Copeland2004}) and deal with them in a different way. Firm believers in the obsolescence of the CTT such as Goldin and Wegner \cite{GoldinWegner2005} agree with this author and others that the usual notion of computability is in fact mathematical and not physical. Opinions often diverge only as to the adequacy of the essentially mathematical definition.

Zeno Machines (and related devices) do not, by the classical definition, \emph{compute}.  Neither does, the reader will probably agree, an instance of Brownian motion (so ubiquitous and beloved of engineering, physics and finance) in any intuitive sense \textit{compute} non-recursive numbers although it certainly produces them in what we regard as a physically realisable process. 
One customarily assumes that finite (but unbounded) time and space resources are used by a computational device and that clearly defined states of the apparatus exist at each point of the computation (and afterwards, in the case of a halting computation). All of this is true for a Turing as well as of a quantum gate computation but not of Zeno machines and of the other metaphors for hypercomputation. 
In \cite{Mostowski2005} Mostowski has proved quite elegantly that under some general and weak assumptions about avoiding an actual infinity, the intuitive notion of decidability coincides with Turing-machine decidability. 
It is the author's opinion that the apparent power of these kinds of hypercomputation is simply due to the breaking of a ``finite but unbounded'' principle which is---as in Mostowski's characterisation---fundamental to the idea of computation. 

In summary: the solution to Hilbert's tenth problem has, after all, shown that the class of recursively enumerable sets 
(sets that constitute the range of a function computed by a Turing machine) is identical to the class of Diophantine sets. The Diophantine sets form such a natural class in mathematics that it seems ill-advised to push the definition of computability, which is primarily a mathematical and not a physical notion, beyond that stated in the Church-Turing Thesis (CTT). 
Based on the conviction that (i)  there is no pressing reason to want to make a computer out of every thing in the universe; 
(ii) the halting problem is inescapable for Chatrapur classes of machines; and (iii) Turing computability is in fact an intuitive and simple notion, this author has found no compelling reason for an immediate redefinition of what we mean by \textit{computable}.

\subsection*{Acknowledgement}

The author wishes to thank the editor and the two anonymous referees at the journal \textit{Theoretical Computer Science} for many helpful comments and references.


\begin{thebibliography}{10}
\expandafter\ifx\csname url\endcsname\relax
  \def\url#1{\texttt{#1}}\fi
\expandafter\ifx\csname urlprefix\endcsname\relax\def\urlprefix{URL }\fi

\bibitem{HodgesOnCopeland}
A.~Hodges, The professors and the brainstorms.
\newline\urlprefix\url{http://www.turing.org.uk/philosophy/sciam.html}

\bibitem{church1936}
A.~Church, An unsolvable problem of elementary number theory, American J. Math
  58 (1936) 345--363.

\bibitem{turing1935}
A.~Turing, On computable numbers with an application to the
  {E}ntscheidungsproblem, Proceedings of the London Mathematical Society (2) 42
  (1935) 230--265.

\bibitem{matijasevich1970ru}
{\cyr Yu}.~{\cyr Matiyasevich}, {\cyr Diofantovost\cyrsftsn \ perechislimykh
  mnozhestv}, {\cyr Doklady Akademii Nauk SSSR} 191 (2) (1970) 279--282.

\bibitem{stanford-encyc}
B.~J. Copeland, The {C}hurch-{T}uring {T}hesis, in: Stanford Encyclopedia of
  Philosophy, 1997.
\newline\urlprefix\url{http://plato.stanford.edu/entries/church-turing/}

\bibitem{foucheJSL2000}
W.~L. Fouch\'e, Arithmetical representations of {B}rownian motion, The Journal
  of Symbolic Logic 65 (2000) 421--442.

\bibitem{bernstein97quantum}
E.~Bernstein, U.~Vazirani, Quantum complexity theory, SIAM Journal on Computing
  26(5) (1997) 1411--1473.
\newline\urlprefix\url{http://citeseer.nj.nec.com/383790.html}

\bibitem{DavisMyth}
M.~Davis, The {M}yth of {H}ypercomputation, in: \cite{Teuscher}.

\bibitem{Akl2005}
S.~Akl, The {M}yth of {U}niversal {C}omputation, Technical Report. 2005-492,
  School of Computing, Queen's University, Kingston, Ontario (January 2005).
\newline\urlprefix\url{http://www.cs.queensu.ca/home/akl/techreports/universal%
.pdf}

\bibitem{Yao2003}
A.~C.-C. Yao, Classical physics and the {C}hurch--{T}uring {T}hesis, Journal of
  the ACM 50~(1) (2003) 100--105.

\bibitem{TimpsonPhD}
C.~G. Timpson, Quantum {I}nformation {T}heory and {T}he {F}oundations of
  {Q}uantum {M}echanics, Ph.D. thesis, The Queen's College (Oxford) (2004).
\newline\urlprefix\url{http://arxiv.org/abs/quant-ph/0412063}

\bibitem{weyl1949}
H.~Weyl, Philosophy of Mathematics and Natural Science, Princeton University
  Press, Princeton, 1949.

\bibitem{Copeland2002}
B.~Copeland, Accelerating {T}uring {M}achines, Minds and Machines 12~(2) (2002)
  281--300.

\bibitem{Copeland2004}
B.~Copeland, {Hypercomputation: philosophical issues}, Theoretical {C}omputer
  {S}cience 317 (2004) 251--267.

\bibitem{BoolosJeffrey1980}
G.~Boolos, R.~C. Jeffrey, Computability and logic, Cambridge University Press,
  1980.

\bibitem{CaludePaun2004}
C.~S. Calude, G.~P\u{a}un, Bio-steps beyond {T}uring, Biosystems 77~(1--3)
  (2004) 175--194.

\bibitem{svozil1998}
K.~Svozil, The {C}hurch--{T}uring thesis as a guiding principle for physics,
  in: C.~Calude, J.~Casti, M.~Ninneen (Eds.), Unconventional Models of
  Computation, Springer, 1998, pp. 371--385.
\newline\urlprefix\url{http://tph.tuwien.ac.at/~svozil/publ/ct.htm}

\bibitem{Thomson1954}
J.~Thomson, Tasks and {S}uper-{T}asks, Analysis 15 (1954--55) 1--13.

\bibitem{etesi}
G.~Etesi, Note on a reformulation of the strong cosmic censor conjceture based
  on computability (2002).
\newline\urlprefix\url{http://arxiv.org/abs/gr-qc/0207086}

\bibitem{etesinemeti}
G.~Etesi, I.~N\'emeti, Non-{T}uring computations via {M}alament-{H}ogarth
  space-times, International Journal of Theoretical Physics 41 (2002) 341--370.

\bibitem{Putnam1965}
H.~Putnam, Trial and {E}rror {P}redicates and the {S}olution to a {P}roblem of
  {M}ostowski, The Journal of Symbolic Logic 30~(1) (1965) 49--57.

\bibitem{Burgin1983}
M.~Burgin, {I}nductive {T}uring {M}achines with a {M}ultiple {H}ead and
  {K}olmogorov {A}lgorithms, Soviet Mathematics Doklady 29~(2) (1984) 189--193,
  translation from the Russian {\cyr Dokl. Akad. Nauk SSSR, {\sc Tom} 275
  (1984) N02}.

\bibitem{BurginKlinger2004}
M.~Burgin, A.~Klinger, Experience, generations, and limits in machine learning,
  Theoretical Computer Science 317~(1-3) (2004) 71--91.

\bibitem{Hamkins2005}
J.~D. Hamkins, Infinitary {C}omputability with {I}nfinite {T}ime {T}uring
  {M}achines, in: \cite{CiE2005}.

\bibitem{yanofsky}
N.~S. Yanofsky, A universal approach to self-referential paradoxes,
  incompleteness and fixed points, Bulletin of Symbolic Logic 09~(3) (2003)
  362--386.
\newline\urlprefix\url{http://arxiv.org/abs/math/0305282}

\bibitem{Cotogno2003}
P.~Cotogno, Hypercomputation and the {P}hysical {C}hurch-{T}uring {T}hesis,
  British {J}ournal for {P}hilosophy of {S}cience 54 (2003) 181--223.

\bibitem{OrdKieu2005}
T.~Ord, T.~D. Kieu, The {D}iagonal {M}ethod and {H}ypercomputation, British
  {J}ournal for the {P}hilosophy of {S}cience 56~(1) (2005) 147--156.
\newline\urlprefix\url{http://www.illc.uva.nl/Publications/ResearchReports/X-2%
005-01.text.pdf}

\bibitem{pourel1982}
M.~B. Pour-El, I.~Richards, Simulating {P}hysics with {C}omputers,
  International Journal of Theoretical Physics 21 (6/7) (1982) 553--555.

\bibitem{feynman1982}
R.~P. Feynman, Simulating {P}hysics with {C}omputers, International Journal of
  Theoretical Physics 21 (6/7) (1982) 467--488.

\bibitem{Manin1980}
{\cyr Yu.I. Manin}, {\cyr Vychislemoe i nevychislimoe}, {\cyr Sovet{s}koe
  Radio}, Moscow, 1980.

\bibitem{potgieter2004}
P.~H. Potgieter, Die voorgeskiedenis van kwantumberekening, Suid-Afrikaanse
  Tydskrif vir Natuurwetenskap en Tegnologie 23 (1/2) (2004) 1--5.
\newline\urlprefix\url{http://arxiv.org/abs/cs/0402037}

\bibitem{potgieteronmanin}
P.~H. Potgieter, Manin's 1980 remarks on quantum computers.
\newline\urlprefix\url{http://www.kolmogorov.net/php/manin.pdf}

\bibitem{KitaevEA}
A.~Y. Kitaev, A.~H. Shen, M.~N. Vyalyi, Classical and {Q}uantum {C}omputation,
  American Mathematical Society, Providence, Rhode Island, 2002.

\bibitem{deutsch85quantum}
D.~Deutsch, Quantum theory, the {C}hurch-{T}uring principle and the universal
  quantum computer, Proceedings of the Royal Society of London Ser.~A A400
  (1985) 97--117.
\newline\urlprefix\url{http://citeseer.nj.nec.com/deutsch85quantum.html}

\bibitem{Shi2002}
Y.~Shi, Remarks on universal quantum computer, Physics Letters A 293~(5--6)
  (2002) 277--282.

\bibitem{Myers1997}
J.~M. Myers, Can a {U}niversal {Q}uantum {C}omputer {B}e {F}ully {Q}uantum?,
  Physical Review Letters 78 (1997) 1823–1824.

\bibitem{simon94power}
D.~R. Simon, On the {P}ower of {Q}uantum {C}omputation, in: Proceedings of the
  35th Annual Symposium on Foundations of Computer Science, Institute of
  Electrical and Electronic Engineers Computer Society Press, Los Alamitos, CA,
  1994, pp. 116--123.
\newline\urlprefix\url{http://citeseer.nj.nec.com/simon94power.html}

\bibitem{shor-introduction}
P.~Shor, Introduction to quantum algorithms.
\newline\urlprefix\url{http://xxx.lanl.gov/abs/quant-ph/0005003}

\bibitem{grover96fast}
L.~K. Grover, A fast quantum mechanical algorithm for database search.
\newline\urlprefix\url{http://citeseer.nj.nec.com/grover96fast.html}

\bibitem{calude2002}
C.~S. Calude, B.~Pavlov, Coins, {Q}uantum {M}easurements, and {T}uring's
  {B}arrier, Quantum Information Processing 1 (2002) 107--127.

\bibitem{Kieu2004}
T.~D. Kieu, Hypercomputation with quantum adiabatic processes, Theoretical
  {C}omputer {S}cience 317~(1--3) (2004) 93--104.

\bibitem{Tsirelson2001}
B.~Tsirelson, The quantum algorithm of {K}ieu does not solve the {H}ilbert's
  tenth problem, quant-ph/0111009 (2001).
\newline\urlprefix\url{http://arxiv.org/abs/quant-ph/0111009/}

\bibitem{Kieu2005}
T.~D. Kieu, Reply to ``{T}he quantum algorithm of {K}ieu does not solve the
  {H}ilberts tenth problem'', quant-ph/0111020 (2005).
\newline\urlprefix\url{http://arxiv.org/pdf/quant-ph/0111020}

\bibitem{AharonovEA2004}
D.~Aharonov, W.~van Dam, J.~Kempe, Z.~Landau, S.~Lloyd, O.~Regev, Adiabatic
  {Q}uantum {C}omputation is {E}quivalent to {S}tandard {Q}uantum
  {C}omputation, in: FOCS '04: Proceedings of the 45th Annual IEEE Symposium on
  Foundations of Computer Science (FOCS'04), IEEE Computer Society, Washington,
  DC, USA, 2004, pp. 42--51.

\bibitem{reif98alternative}
J.~H. Reif, {A}lternative {C}omputational {M}odels: {A} {C}omparison of
  {B}iomolecular and {Q}uantum {C}omputation, FSTTCS: Foundations of Software
  Technology and Theoretical Computer Science 18.
\newline\urlprefix\url{http://www.cs.duke.edu/~reif/paper/paper.html/altcomp.p%
s}

\bibitem{reif-emergence}
J.~H. Reif, The {E}mergence of the {D}iscipline of {B}iomolecular {C}omputation
  in the {U}{S} (2002).
\newline\urlprefix\url{http://citeseer.nj.nec.com/reif02emergence.html}

\bibitem{longo2002}
G.~Longo, Lo spazio, i fondamenti della matematica e la resistibile ascesa
  della metafora: il cervello \`e un calcolatore digitale, in: M.~B. Califano
  (Ed.), L'uomo e le macchine, Leo S. Olschki, Firenze, 2002.
\newline\urlprefix\url{ftp://ftp.di.ens.fr/pub/users/longo/PhilosophyAndCognit%
ion/metafora-resist.pdf}

\bibitem{GoldinWegner2005}
D.~Goldin, P.~Wegner, The {C}hurch-{T}uring {T}hesis: {B}reaking the {M}yth,
  in: \cite{CiE2005}.

\bibitem{Mostowski2005}
M.~Mostowski, Potential {I}nfinity and the {C}hurch {T}hesis, in:
  \cite{CiE2005math}.

\bibitem{Teuscher}
C.~Teuscher (Ed.), Alan Turing: Life and Legacy of a Great Thinker, Springer,
  2004.

\bibitem{CiE2005}
S.~B. Cooper, B.~L{\"o}we, L.~Torenvliet (Eds.), New Computational Paradigms,
  Vol. 3526 of Lecture {N}otes in {C}omputer {S}cience, Springer, 2005.

\bibitem{CiE2005math}
S.~B. Cooper, B.~L{\"o}we, L.~Torenvliet (Eds.), CiE 2005: New Computational
  Paradigms, Technical Notes (X) Series, Institute for {L}ogic, {L}anguage and
  {C}omputation, University of Amsterdam, 2005.
\newline\urlprefix\url{http://www.illc.uva.nl/Publications/ResearchReports/X-2%
005-01.text.pdf}

\end{thebibliography}

\end{document}